\documentclass[preprint,pra,showkeys,showpacs,altaffilletter,superscriptaddress,groupedaddress]{revtex4}

\usepackage{amsmath}
\usepackage{amsfonts}
\usepackage{latexsym,amssymb}
\usepackage{graphicx}
\usepackage{bm}
\newcommand{\ket}[1]{\ensuremath{ \left| #1 \right\rangle} }
\begin{document}

\title{Quantum number dimensional scaling analysis for\\
excited states of multielectron atoms}

\author{Robert K. Murawski}
\email{rmurawski@physics.tamu.edu}
\affiliation{Institute for Quantum Studies and Department of Physics, Texas A\,\&\,M University, 77843-4242}
\author{Anatoly A. Svidzinsky}
\affiliation{Institute for Quantum Studies and Department of Physics, Texas A\,\&\,M University, 77843-4242}
\date{\today}

\begin{abstract}
A new dimensional scaling method for the calculation of excited states of
multielectron atoms is introduced. By including the principle and orbital
quantum numbers in the dimension parameter, we obtain an energy expression
for excited states including high angular momentum states.  The method is tested
on He, Li, and Be.  We obtain good agreement with more orthodox quantum mechanical
treatments even in the zeroth order.
\end{abstract}

\maketitle

\section{Introduction}
The dimensional scaling (D-scaling) method originally developed by
Witten \cite{ew80} and Herschbach \cite{dh86} is now a well
established technique in atomic physics \cite{dh93}.  Substantial progress
has been made in improving the accuracy of the technique and extending it to
excited states.  Most notable are the approaches of Goodson and Herschbach \cite{dg87,dg92,dg93} and the
dimensional perturbation theory (DPT) developed by Carzoli, Dunn, and Watson \cite{jc99,md99}.
In the present work, an alternative approach to generalized D-scaling is developed which does not
require calculations of high order $1/D$ corrections in order to obtain
high-angular momentum states unlike, for example, the Pad\'e approximate
used by Goodson which required 20 orders of $1/D$.  Our approach is simple and yet
accurate. It predicts excited states of helium (the canonical multi-electron problem)
and is easily extended to N electron atoms.  We demonstrate the straight forward extension
by analyzing lithium and beryllium.

\section {Two approaches to D-scaling}
In this section, we will review the traditional D-scaling
approach for helium and introduce our alternative quantum number dimensional
formulation.
\subsection{Traditional D-scaling for helium}

The Hamiltonian for helium, in atomic units, is given by
\begin{equation}
\label{Ham}\widehat{H}=-\frac 12(\nabla _1^2+\nabla _2^2)+V(r_1,r_2,\theta),
\end{equation}
where
\begin{equation}
\label{Coul}
V(r_1,r_2,\theta)=
-\frac Z{r_1}-\frac Z{r_2}+\frac 1{\sqrt{r_1^2+r_2^2-2{r_1}{r_2%
}\cos {\theta }}}
\end{equation}
is the Coulomb potential energy, $\theta$ is the angle between the
electron radii vectors, and $Z$ is the nuclear charge.
The Laplacian operator, for $\bf{S}$ states, can be written in D-dimensions as
\begin{equation}\label{lap}
\nabla_1^2 + \nabla_2^2 = K_{D-1}(r_1)+K_{D-1}(r_2)+\left( \frac{1}{r_1^2}+\frac{1}{r_2^2} \right)L^2_{D-1}
\end{equation}
where
\begin{subequations}\label{kl2}
\begin{eqnarray}
K_{D-1}(r)= \frac{1}{r^{D-1}}\frac{\partial}{\partial{r}}r^{D-1}\frac{\partial}{\partial{r}}\\
L^2_{D-1}=\frac{1}{\sin^{D-2}{\theta}}\frac{\partial}{\partial\theta}\sin^{D-2}{\theta}\frac{\partial}{\partial\theta}.
\end{eqnarray}
\end{subequations}
The probability distribution function in D-dimensions is defined to be
${|\Phi_D|}^2 = J_D {|\Psi_D|}^2$ where the D-dimensional Jacobian is given by
\begin{equation}\label{jac}
J_D = (r_1r_2)^{D-1}\sin^{D-2}{\theta}
\end{equation}
and $\Psi_D$ is a solution to (\ref{Ham}) with eigenvalue $E$.  Letting
(\ref{lap}) act on $\Psi_D=J_D^{-\frac{1}{2}}\Phi_D$ and transforming parameters as
\begin{subequations}\label{xfrms}
\begin{eqnarray}
r_i \rightarrow \frac{(D-1)^2}{4}r_i\\
E \rightarrow \frac{4}{(D-1)^2}E
\end{eqnarray}
\end{subequations}
we obtain the $D \rightarrow \infty$ limit of the energy
\begin{equation}\label{energy}
E_{\infty} = \frac{1}{2}\left( \frac{1}{r_1^2}+\frac{1}{r_2^2}\right)\frac{1}{\sin^2\theta}+V(r_1,r_2,\theta).
\end{equation}
When this energy expression is minimized with respect to the parameters $(r_1, r_2, \theta)$, an estimate of
the ground state ($1s^2$) energy of helium can be obtained which is $5\%$ accurate.  The accuracy can be
improved by the $1/D$ expansion. The limitation of this approach is that expression (\ref{energy}) cannot
produce excited $\bf{S}$ states nor higher-angular momentum states (states with $L>0$) in its current form
without resorting to Langmuir vibrations and high order $1/D$ corrections.

\subsection{Present D-scaling for helium}

Motivated by the approach of the previous section
and to avoid high order dimensional perturbation expansions, we invented a
generalized Laplacian of the form
\begin{equation}\label{lap2}
\nabla_1^2 + \nabla_2^2 = K_{\gamma_1}(r_1)+K_{\gamma_2}(r_2)+\left( \frac{1}{r_1^2}+\frac{1}{r_2^2} \right)L^2_\alpha
\end{equation}
where $\gamma_1, \gamma_2$, and $\alpha$ are three different parameters which we also assume will
tend toward $\infty$.  Without
loss of generality, we then factor our parameters into a finite
integer part multiplied by a dimensional term given by
\begin{subequations}\label{para}
\begin{eqnarray}
\gamma_1 =  n_1(D-1)\\
\gamma_2 =  n_2(D-1)\\
\alpha   =  L(D-2).
\end{eqnarray}
\end{subequations}
Thus, $\gamma_i$ and $\alpha$ will tend toward $\infty$ faster
than $D$ provided n$_i$ and $L$ are finite positive integers greater
than one.  By introducing the parameters n$_i$ and $L$,
into the usual D dimensional Laplacian, we will have
additional degrees of freedom in the final energy expression.  This parameterization
can be interpreted as being the cardinality of a space larger
than D dimensions which includes the energy levels of the physical system.  We will refer to
this space as quantum number dimensional space.  If we need to recover the usual three
dimensional Laplacian, we can add a second term to Eq. \ref{para} given by
\begin{subequations}\label{para2}
\begin{eqnarray}
\gamma_1 =  n_1(D-1)+6(1-n_1)/D\\
\gamma_2 =  n_2(D-1)+6(1-n_2)/D\\
\alpha   =  L(D-2)+3(1-L)/D.
\end{eqnarray}
\end{subequations}
This parameterization allows us to recover the usual
three dimensional Schr\"odinger equation when $D=3$.  But for large values of D, system
\ref{para} and system \ref{para2} are the same.
In $\gamma-\alpha$ space, the corresponding Jacobian factor now reads
\begin{equation}\label{jac2}
J_{\gamma-\alpha}=r_1^{\gamma_1}r_2^{\gamma_2}\sin^\alpha{\theta}
\end{equation}
which is chosen to provide consistency with the traditional approach.
If we now solve $\hat{H}\Psi_{\gamma-\alpha}=E\Psi_{\gamma-\alpha}$ with definitions (\ref{lap2}), (\ref{para}), and
the scaling transformation (\ref{xfrms}) we obtain the final energy expression in the $D \rightarrow \infty$ as
\begin{equation}\label{energy2}
E=\frac{1}{2}\left(\frac{n_1^2}{r_1^2}+\frac{n_2^2}{r_2^2}\right) +
\frac{L^2}{2}\left(\frac{1}{r_1^2}+\frac{1}{r_2^2}\right)
\cot^2\theta+V(r_1,r_2,\theta).
\end{equation}

The derivation of this expression is given in the Appendix A. 
Additionally, in the Appendix B we show how to correct
the energy with the first $1/D$ correction.
Equation (\ref{energy2}) will allow us to obtain
atomic excited states by associating n$_i$ as the principle quantum number for electron $i$ and $L$ as
the sum of the individual hydrogenic assignment orbital quantum numbers i.e. $L = \ell_1 + \ell_2$.
Thus, our $L$ is the maximum in the set of possible total orbital angular momenta
$\bf{L} \in \{|\ell_1 - \ell_2|,\ldots,|\ell_1 + \ell_2|\}$.

\section{Ground state and excited S states}
The excited $S$ states of helium can easily be found by setting $L=0$ in expression (\ref{energy2}),
we find
\begin{equation}\label{energy2s}
E=\frac{1}{2}\left(\frac{n_1^2}{r_1^2}+\frac{n_2^2}{r_2^2}\right) +V(r_1,r_2,\theta).
\end{equation}
It is interesting to note that this energy expression is identical to what one would obtain
for a Bohr model of helium using the Bohr space quantization condition for the kinetic energy
operators i.e. $2\pi r = n\lambda$ where $\lambda$ is the de Broglie wavelength $\lambda = h/p$.
Bohr space quantization was also used by Greenspan \cite{dg86} to obtain an energy expression for
the excited states of helium but with a potential function that was parametrized.  More recently, the
Bohr model has been successfully applied to molecules
\cite{aspnas05,asprl05,Svid06,Chen05}.
Here, we take the electron-nuclear and electron-electron interaction to have their usual
Coulombic form.  The results for excited $\bf{S}$ states are presented
in Table  \ref{tab_leq0}.
\begin{table}
\caption{\label{tab_leq0}
Energy (in atomic units) of the He atom obtained using Eq. (\ref{energy2s})
for stats with $L=0$.
The final column is the percent error between our energy and previously published
results.}
\begin{ruledtabular}
\begin{tabular}{ccccc}
$\ket{n_1n_2L}$ &Assignment &-E (Present.)
&-E (Ref.)\footnotemark[1] & $\Delta (\%)$ \\
\hline
$\ket{110}$ & $1s^2$ $^1S$ & 3.0625 & 2.9037 & 5.40\\
$\ket{120}$ & $1s2s$ $^1S$ & 2.1595 & 2.1460 & 0.60\\
$\ket{130}$ & $1s3s$ $^1S$ & 2.0621 & 2.0613 & 0.03\\
$\ket{140}$ & $1s4s$ $^1S$ & 2.0333 & 2.0336 & 0.02\\
\end{tabular}
\end{ruledtabular}
\footnotetext[1]{Complex rotation technique \cite{ab95}}
\end{table}
We see that a dominant feature is the increase in accuracy with larger
quantum number $n_2$.  This can easily be explained in terms of Bohr
correspondence.  Physically, as $n_2 > n_1(=1)$ the corresponding radii will
be very different since $r_i \sim n_i^2$.  Thus, electron two sees a screened
nucleus of charge $Z=1$ and therefore it is a hydrogenic problem, which is
exactly soluble in the Bohr picture.  
\section{Excited P and D states} 
Our goal here is to stay within the geometric configuration picture originally
established by Hershbach and co-workers.  Other groups have also made
substantial contributions (see \cite{dh93} and references therein) to use
D-scaling for these states.  We present here, for the first time, a true Lewis
structure picture for helium excited states thus providing an example of a
classical analogue for $P$ states.
\begin{table}
\caption{\label{tab:lgt0}
Energy of the He atom for states with $L>0$.}
\begin{ruledtabular}
\begin{tabular}{ccccc}
$\ket{n_1n_2L}$ &Assignment &-E (Present.) & -E (Ref.)
& $\Delta (\%)$ \\
\hline
$\ket{121}$ & $1s2p$ $^1P$ & 2.12710 & 2.12380\footnotemark[2] & 0.20\\
$\ket{131}$ & $1s3p$ $^1P$ & 2.05573 & 2.05514\footnotemark[2] & 0.020\\
$\ket{132}$ & $1s3d$ $^1D$ & 2.05572 & 2.05562\footnotemark[2] & 0.004\\
$\ket{221}$ & $2s2p$ $^1P$ & 0.69720 & 0.69310\footnotemark[3] & 0.60\\
$\ket{222}$ & $2p^2$ $^1D$ & 0.68440 & 0.68641\footnotemark[3] & 0.29\\
$\ket{231}$ & $2s3p$ $^1P$ & 0.55990 & 0.56384\footnotemark[3] & 0.70\\
\end{tabular}
\end{ruledtabular}
\footnotetext[2]{Reference \cite{ed64}}
\footnotetext[3]{Reference \cite{lh94}}
\end{table}
Physically, the $P$ and $D$ states are just special configurations of the two electron
system in hyperspherical coordinates. Table \ref{tab:lgt0} compares the energy of the He
atom for some states with $L>0$ obtained in the present analysis with
the accurate values known in the literature.
Table \ref{tab:coords} shows the configurations
of the electrons for some representative states.  We have found that pure $S$ states admit
an angle of $180^\circ$ whereas states with $P$ character and higher-angular momenta
typically lie in the range of $90^\circ < \theta < 180^\circ$.
\begin{table}
\caption{\label{tab:coords}Positions of the electrons on the hypersphere for
certain representative states}
\begin{ruledtabular}
\begin{tabular}{cccc}
 $\ket{n_1n_2L}$ & $r_1$ & $r_2$ & $\theta$ \\
\hline
$\ket{121}$ & $0.5005$ & $3.8981$ & $90.4526^\circ$\\
$\ket{132}$ & $0.5000$ &  $8.9579$ & $90.0221^\circ$\\
$\ket{222}$ & $2.4278$ & $2.4278$ & $95.3006^\circ$\\
$\ket{231}$ & $2.0249$ & $7.9728$ & $95.8483^\circ$\\
\end{tabular}
\end{ruledtabular}
\end{table}
We cannot, with this method, distinguish between singlet and triplet states of helium nor can
we determine the correct spectroscopic assignment of $\bf{L}$.  We note that in general, our
method finds a value somewhere near the average of the various $\bf{L}$ states.
For example, the $2p^2\medskip$ $^1S^e$ state energy
is $\approx -0.62 $ a.u. whereas for the $2p^2\medskip$ $^3P^e$ state $E\approx
-0.71$ a.u. and the average of the two yields $-0.67$ a.u. which is much closer
to our result of $-0.68$ a.u.  However, our calculated energy is only 0.29\% different
from the $^1D$ state.  This suggests that our $L$ is indeed the sum of the
individual $\ell_i$ s.

Our advantage over the dimensional perturbation treatment is twofold.  Firstly, in
order to calculate the excited states of helium they needed to invoke harmonic
oscillator wavefunctions as an ansatz to the problem. The various excited
modes of the normal coordinates corresponded to the hydrogenic excited states
of helium.  Their harmonic oscillator quantum numbers where associated with the
radial and orbital nodes of the hydrogenic wavefunctions (see for example \cite{jl86}). 
In order to calculate with this method, many orders of $1/D^n$ (in the $1/D$ expansion)
where required. Furthermore, it was seen that the $1/D$ series was asymptotic
and required further re-summation via Pad\'e approximates.  The second drawback
was the fact that it could only consider states where $\ell_1 = \ell_2$.
Thus, the $1s2p$ state is not accounted for in their treatment.

The states with electron $1$ in an s-orbit and electron $2$ in a higher angular
momentum state seem to work the best.  We can correct the errors by considering
small vibrations about the ``frozen'' configuration positions.  The derivation of
the first $1/D$ correction and its application are provided in the
Appendix B for the excited states of helium.

\section{Generalization to the N-electron system}
The present approach admits a simple generalization to the N electron atom with
nuclear charge $Z$.  We can describe the $N$ electron system by including
all pairwise interactions.  The problem will thus involve $(1/2)N(N+1)$ parameters.
Let
\begin{equation}\label{ki}
K_{\gamma_i}= \frac{1}{r^{\gamma_i}}\frac{\partial}{\partial r_i}\left(r_i^{\gamma_i}\frac{\partial}{\partial r_i}\right)
\end{equation}
as before and
\begin{equation}\label{l2i}
L^2_{\alpha_{ij}}=\frac{1}{\sin^{\alpha_{ij}}\theta_{ij}}\frac{\partial}{\partial\theta_{ij}}
\left({\sin^{\alpha_{ij}}\theta_{ij}}\frac{\partial}{\partial\theta_{ij}}\right)
\end{equation}
then
\begin{eqnarray}\label{hn}
\hat{H}=-\frac{1}{2}\left[\sum^{N}_{i=1}K_{\gamma_i} + \sum^{N}_{i<j}
\left(\frac{1}{r_i^2}+\frac{1}{r_j^2}\right)L^2_{\alpha_{ij}}\right]\\
-Z\sum^N_{i=1}\frac{1}{r_i}+\sum^{N}_{i<j}
(r_i^2+r_j^2-2r_ir_j\cos\theta_{ij})^{-{\frac{1}{2}}}. \nonumber
\end{eqnarray}
Here $\gamma_i$ is as before and $\alpha_{ij}=L_{ij}(D-2)$ with $L_{ij}=\ell_i + \ell_j$.
Next we transform the $N$ particle wave-function as
\begin{equation}\label{psin}
{\ket{\Psi}}=\prod^N_{i = 1}r_i^{-\frac{\gamma_i}{2}}
\prod^N_{i<j}sin^{-\frac{\alpha_{ij}}{2}}\theta_{ij}\ket{\Phi}.
\end{equation}
Transforming
${\ket{\Psi}}$ in this way makes the analysis a natural continuation of the helium problem with the exception
that the angular momentum are multiply counted. Then in the large-D limit
the energy function to be minimized is
$$
E(n_1 \ell_1 n_2\ell_2 \ldots n_N\ell_N)=
\frac{1}{2}\sum^{N}_{i=1}\frac{n_i^2}{r_i^2}+
$$
\begin{equation}
+\frac{1}{2}\left(\frac{1}{N-1}\right)\sum_{i<j}L^2_{ij}\left(\frac{1}{r_i^2}+
\frac{1}{r_j^2}\right)\cot^2\theta_{ij}+
V(\vec{r},\vec{\theta}),
\end{equation}
where
\begin{equation}
V(\vec{r},\vec{\theta})=-Z\sum\frac{1}{r_i}+
\sum_{i<j}(r_i^2+r_j^2-2r_ir_j\cos\theta_{ij})^{-\frac{1}{2}}.
\end{equation}
The factor of $1/(N-1)$ accounts for the multiple counting of each $\ell_i$.
For the case of the ground state of lithium $(Z=3,N=3)$ in the state $1s^22s$
we obtain $E=-7.7468$ which is $3.5\%$ different from the true ground state of
$-7.4780$.  However, the accuracy of higher states (including $\ell_i >0 $) is
typically much better (see table \ref{tab:lithium}).
\begin{table}
\caption{\label{tab:lithium}
Results for
various excited states of Lithium} \begin{ruledtabular} \begin{tabular}{ccccc}
$\ket{n_1\ell_1n_2\ell_2n_3\ell_3}$ &Assignment
&-E (Present.) & -E (Ref.) & $\Delta (\%)$ \\
\hline
$\ket{101021}$ & $1s^22p$ $^2P$ & 7.6897 & 7.4101\footnotemark[4] & 3.70\\
$\ket{102021}$ & $1s2s2p$ $^4P$ & 5.2715 & 5.4375\footnotemark[5] & 3.10\\
$\ket{102121}$ & $1s2p^2$ $^1D$ & 5.1982 & 5.2335\footnotemark[6] & 0.70\\
$\ket{102130}$ & $1s2p3s$ $^1P$ & 5.0744 & 5.0820\footnotemark[7] & 0.15\\
$\ket{102131}$ & $1s2p3p$ $^1P$ & 5.0668 & 5.0651\footnotemark[6] & 0.03\\
$\ket{102141}$ & $1s2p4p$ $^3S$ & 5.0395 & 5.0311\footnotemark[6] & 0.16\\
$\ket{202032}$ & $2s^23d$ $^3P$ & 1.9496 & 1.9690\footnotemark[8] & 1.00\\
$\ket{212132}$ & $2p^23d$ $^3P$ & 1.8192 & 1.8246\footnotemark[8] & 0.30\\
\end{tabular}
\end{ruledtabular}
\footnotetext[4]{Reference \cite{cc68}}
\footnotetext[5]{Reference \cite{jg65}}
\footnotetext[6]{Reference \cite{kc81}}
\footnotetext[7]{Reference \cite{jc70}}
\footnotetext[8]{Reference \cite{hz99}}
\end{table}
Additionally, we tested the method on beryllium and had some success.  In this study, we considered
states with the core electrons fixed in the $1s^2$ orbits.  We found an error with respect to published
values of 2.6 percent.  To eliminate this systematic error, we subtracted off the ground state error
from each of the excited states and found very good agreement with published results.  Shown in the
table \ref{tab:belgt0} are some calculated excited states of beryllium which are compared to
other published results.
\begin{table}
\caption{\label{tab:belgt0}Ground state and various excited states of beryllium.  Core electrons
$1s^2$ are implicit in the notation.  The percent error is wrt the corrected value column.}
\begin{ruledtabular}
\begin{tabular}{cccccc}
 $\ket{n_3\ell_3n_4\ell_4}$ &Assignment &-E(Present.) & $Corrected$ & -E(Ref.)\footnotemark[9] & $\Delta (\%)$ \\
\hline
$\ket{2020}$ & $2s^2$ $^1S$ & 15.0603 & 14.6673 & 14.6673 & 0.00\\
$\ket{2021}$ & $2s2p$ $^1P$ & 14.8939 & 14.5009 & 14.4734 & 0.20\\
$\ket{2030}$ & $2s3s$ $^1S$ & 14.8014 & 14.4084 & 14.4182 & 0.07\\
$\ket{2121}$ & $2p^2$ $^1D$ & 14.7674 & 14.3744 & 14.4079 & 0.23\\
$\ket{2031}$ & $2s3p$ $^1P$ & 14.7693 & 14.3763 & 14.3931 & 0.11\\
$\ket{2032}$ & $2s3d$ $^1D$ & 14.7682 & 14.3752 & 14.3737 & 0.01\\
\end{tabular}
\end{ruledtabular}
\footnotetext[9]{References \cite{fg02}, \cite{lb01}, and \cite{kc93}}
\end{table}

\section{Conclusion}
We have introduced an alternative form of dimensional scaling for the
excited states of atoms.  By treating the principle and orbital quantum numbers
to be innately coupled to the dimension D, we developed an algebraic equation
for multiple excited states of multielectron atoms including states of
high orbital momentum. This method can be extended to include relativistic effects
for nuclei with high Z numbers by including the kinetic-energy mass correction term
of the Breit-Pauli Hamltonian.  

\begin{acknowledgments}
One of the authors (RKM) would like to thank Prof. David Goodson for a
useful discussion.  The authors would like to thank Profs. Marlan Scully and
Dudley Herschbach for useful insights and encouragement.  The
authors would like to thank Texas A \& M and the Robert A. Welsh Foundation for
supporting this work.
\end{acknowledgments}

\appendix

\begin{widetext}
\onecolumngrid

\section{Schr\"odinger equation in ``D dimensions''}

Here we demonstrate the idea of our D-scaling transformation by applying it
to the He atom. The Hamiltonian of He in three dimensions is given by
Eq. (\ref{Ham}). First we perform a continuous
transformation of the Laplacian as follows
\begin{equation}
\nabla _1^2+\nabla _2^2=\frac 1{r_1^{\gamma _1}}\frac \partial {\partial
r_1}r_1^{\gamma _1}\frac \partial {\partial r_1}+\frac 1{r_2^{\gamma
_2}}\frac \partial {\partial r_2}r_2^{\gamma _2}\frac \partial {\partial
r_2}
+\left( \frac 1{r_1^2}+\frac 1{r_2^2}\right) \frac 1{\sin
^\alpha \theta }\frac \partial {\partial \theta }\sin ^\alpha \theta \frac
\partial {\partial \theta },
\end{equation}
where $\gamma _1$, $\gamma _2$ and $\alpha $ are given by Eqs. (\ref{para}).
The wave
function, coordinates and the energy transform according to
\begin{equation}
\Psi =\left[ r_1^{\gamma _1}r_2^{\gamma _2}\sin ^\alpha {\theta }%
\right] ^{-1/2}\Phi ,\quad r_i\rightarrow \frac{(D-1)^2}4r_i,\quad
E\rightarrow \frac 4{(D-1)^2}E.
\end{equation}
The radial part of the Laplacian acting on $\Psi $ yields
\begin{equation}
\label{radial}\frac 1{r^\gamma }\frac \partial {\partial r}\left( r^\gamma
\frac{\partial \Psi }{\partial r}\right) =r^{-\frac \gamma 2}\sin ^{-\frac
\alpha 2}\theta \times \left[ \frac{\partial ^2}{\partial r^2}-\left( \frac
\gamma 2\right) \left( \frac{\gamma -2}2\right) \frac 1{r^2}\right] \Phi ,
\end{equation}
while the angular part leads to
\begin{equation}
\label{angular}\frac 1{\sin ^\alpha \theta }\frac \partial {\partial \theta
}\left( {\sin ^\alpha \theta }\frac{\partial \Psi }{\partial \theta }\right)
=r^{-\frac \gamma 2}\sin ^{-\frac \alpha 2}\theta \times \left[ \frac{%
\partial ^2}{\partial \theta ^2}+\frac \alpha 2-\left( \frac \alpha 2\right)
\left( \frac{\alpha -2}2\right) \cot ^2\theta \right]
\Phi .
\end{equation}
After the transformation the Schr\"odinger equation reduces to
\begin{equation}
\label{s1}(T+U+V)\Phi =E\Phi ,
\end{equation}
where
\begin{equation}
\label{tee}T=-\frac 2{(D-1)^2}\left[ \frac{\partial ^2}{\partial r_1^2}+%
\frac{\partial ^2}{\partial r_2^2}+\left( \frac 1{r_1^2}+\frac
1{r_2^2}\right) \frac{\partial ^2}{\partial \theta ^2}\right]
\end{equation}
is the kinetic energy term and
\begin{equation}
\label{cent}U=\frac{\gamma _1(\gamma _1-2)}{2(D-1)^2}\frac 1{r_1^2}+\frac{%
\gamma _2(\gamma _2-2)}{2(D-1)^2}\frac 1{r_2^2}+\left( \frac 1{r_1^2}+\frac
1{r_2^2}\right) \left[ \frac{\alpha (\alpha -2)}{2(D-1)^2}\cot ^2\theta
-\frac \alpha {(D-1)^2}\right]
\end{equation}
is the centrifugal potential. In the large-D limit the kinetic energy term
vanishes and Eq. (\ref{s1}) reduces to finding the minimum of the algebraic energy
function (\ref{energy2}).

\section{Calculation of 1/D correction}

Let us take into account terms of the order of $1/D$ in Eq. (\ref{s1}), then
it reduces to
$$
\left\{ -\frac 2{D^2}\left[ \frac{\partial ^2}{\partial r_1^2}+\frac{%
\partial ^2}{\partial r_2^2}+\left( \frac 1{r_1^2}+\frac 1{r_2^2}\right)
\frac{\partial ^2}{\partial \theta ^2}\right] +\frac 12\left( \frac{n_1^2}{%
r_1^2}+\frac{n_2^2}{r_2^2}\right) +\frac{L^2}2\left( \frac 1{r_1^2}+\frac
1{r_2^2}\right) \cot ^2\theta +V-\right.
$$
\begin{equation}
\label{b1}\left. -\frac 1D\left[ \frac{n_1}{r_1^2}+\frac{n_2}{r_2^2}+\left(
\frac 1{r_1^2}+\frac 1{r_2^2}\right) \left( L(L+1)\cot ^2\theta +L\right)
\right] \right\} \Phi =E\Phi
\end{equation}
We decompose the effective potential $U+V$ in Eq. (\ref{b1}) near the
minimum $(r_{10},r_{20},\theta _0)$ and leave only terms quadratic in
displacement from this point. As a result, we obtain%
$$
\left\{ E_\infty -\frac 1D\left[ \frac{n_1}{r_{10}^2}+\frac{n_2}{r_{20}^2}%
+\left( \frac 1{r_{10}^2}+\frac 1{r_{20}^2}\right) \left( L(L+1)\cot
^2\theta _0+L\right) \right] \right\} \Phi -
$$
$$
\frac 2{D^2}\left[ \frac{%
\partial ^2}{\partial r_1^2}+\frac{\partial ^2}{\partial r_2^2}+\left( \frac
1{r_{10}^2}+\frac 1{r_{20}^2}\right) \frac{\partial ^2}{\partial \theta ^2}%
\right] \Phi +
$$
\begin{equation}
\label{b2}[\beta _{11}\Delta r_1^2+\beta _{22}\Delta r_2^2+\beta _{33}\Delta
\theta ^2+2\beta _{12}\Delta r_1\Delta r_2+2\beta _{13}\Delta r_1\Delta
\theta +2\beta _{23}\Delta r_2\Delta \theta ]\Phi =E\Phi ,
\end{equation}
where $\beta _{ij}$ are coefficients of the Taylor expansion of the function
$U+V$ at the minimum. Eq. (\ref{b2}) describes three coupled one dimension
harmonic oscillators. To make the oscillator masses equal we rescale $\Delta
\theta $ as $\Delta \theta =\Delta \tilde \theta /r_0$, where $1/r_0=\sqrt{%
1/r_{10}^2+1/r_{20}^2}$. Then Eq. (\ref{b2}) yields%
$$
\left\{ E_\infty -\frac 1D\left[ \frac{n_1}{r_{10}^2}+\frac{n_2}{r_{20}^2}%
+\left( \frac 1{r_{10}^2}+\frac 1{r_{20}^2}\right) \left( L(L+1)\cot
^2\theta _0+L\right) \right] \right\} \Phi -\frac 2{D^2}\left[ \frac{%
\partial ^2}{\partial r_1^2}+\frac{\partial ^2}{\partial r_2^2}+\frac{%
\partial ^2}{\partial \tilde \theta ^2}\right] \Phi +
$$
\begin{equation}
\label{b3}[\beta _{11}\Delta r_1^2+\beta _{22}\Delta r_2^2+\beta _{33}\Delta
\tilde \theta ^2/r_0^2+2\beta _{12}\Delta r_1\Delta r_2+2\beta _{13}\Delta
r_1\Delta \tilde \theta /r_0+2\beta _{23}\Delta r_2\Delta \tilde \theta
/r_0]\Phi =E\Phi .
\end{equation}
The problem is reduced to determining eigenvalues $\Lambda _1$, $\Lambda _2$%
, $\Lambda _3$ of the symmetric matrix:
$$
\left(
\begin{array}{ccc}
\beta _{11} & \beta _{12} & \beta _{13}/r_0 \\
\beta _{12} & \beta _{22} & \beta _{23}/r_0 \\
\beta _{13}/r_0 & \beta _{23}/r_0 & \beta _{33}/r_0^2
\end{array}
\right)
$$
The energy including the $1/D$ correction is then given by%
$$
E=\frac 4{(D-1)^2}\left\{ E_\infty -\frac 1D\left[ \frac{n_1}{r_{10}^2}+%
\frac{n_2}{r_{20}^2}+\left( \frac 1{r_{10}^2}+\frac 1{r_{20}^2}\right)
\left( L(L+1)\cot ^2\theta _0+L\right) \right] +\right.
$$
\begin{equation}
\label{b4}\left. \frac{\sqrt{2}}D\left[ \sqrt{\Lambda _1}+\sqrt{\Lambda _2}+%
\sqrt{\Lambda _3}\right] \right\} .
\end{equation}
\end{widetext}
We have applied Eq.(\ref{b4}) to correct the energy we found for the $1s^2$ state from 
table \ref{tab_leq0} and the $2s3p$ state from \ref{tab:lgt0}. Shown in table \ref{tab:correct}
is a comparison between the zeroth order and first 1/D correction for these states with 
respect to the more exact values.
\begin{table}
\caption{\label{tab:correct}Improvement after the first 1/D correction.}
\begin{ruledtabular}
\begin{tabular}{ccc}
 State & Zeroth order & First order \\
\hline
$1s^2$ $^1S$ & 5.40\% & 1.67\% \\
$2s3p$ $^1P$ & 0.70\% & 0.54\% \\
\end{tabular}
\end{ruledtabular}
\end{table}
We have noticed that it is possible to correct the energy with the 1/D expansion to
first order but only if the zeroth order error is significant. For states where the
zeroth order gave very good accuracy (ie < 1\%) we do not expect much improvement from the first 1/D
correction. It was noted in reference \cite{dh93} that 30
terms of the 1/D expansion leads to a 9 decimal place accuracy for the ground state
energy of helium. Thus, the first order term should only correct the most significant figure.

\end{document}